\documentclass[aps,prb,superscriptaddress,amsmath,eqsecnum,showkeys]{revtex4}

   \usepackage{bm}
\usepackage{graphicx,amssymb}
\usepackage{bm}
\usepackage{amsmath}

\newcommand{\one}{{(1)}}

\newcommand{\beq}{\begin{equation}}
\newcommand{\eeq}{\end{equation}}
\newcommand{\beqa}{\begin{eqnarray}}
\newcommand{\eeqa}{\end{eqnarray}}

\newcommand{\nn}{\nonumber\\}
\def\two{{(2)}}

\def\z{\zeta_0^*}
\begin{document}

\title{Poiseuille flow in a heated granular gas}
\author{Mohamed Tij}
\email{mtij@fsmek.ac.ma}
\affiliation{D\'epartment de Physique, Universit\'e Moulay Isma\"{\i}l, Mekn\`es, Morocco}
\author{Andr\'es Santos}
\email{andres@unex.es}
\homepage{http://www.unex.es/fisteor/andres}
\affiliation{Departamento de F\'{\i}sica, Universidad de Extremadura,
E--06071 Badajoz, Spain}

\date{\today}
\begin{abstract}
The planar Poiseuille flow induced by a constant external field (e.g., gravity) has been the subject of recent interest in the case of molecular gases. One of the predictions from kinetic theory (confirmed  by computer simulations) has been that the temperature profile exhibits a bimodal shape with a local minimum in the middle of the slab surrounded by two symmetric maxima, in contrast to the unimodal shape expected from the Navier--Stokes (NS) equations.  However, from a practical point of view, the interest of this non-Newtonian behavior in molecular gases is rather academic since it requires values of  gravity  extremely higher than the terrestrial one. On the other hand, gravity plays a relevant role in the case of granular gases due to the mesoscopic nature of the grains.
In this paper we consider a dilute gas of inelastic hard spheres enclosed in a slab under the action of gravity along the longitudinal direction. In addition, the gas is subject to a white-noise stochastic force that  mimics the effect of  external vibrations customarily used in experiments to compensate for the collisional cooling.  The system is described by means of a kinetic model of the inelastic Boltzmann equation and its steady-state solution is derived through second order in gravity. This solution differs from the NS description in that the hydrostatic pressure is not uniform, normal stress differences are present, a component of the heat flux normal to the thermal gradient exists, and  the temperature profile  includes a positive quadratic term. As in the elastic case, this new term is responsible for the bimodal shape of the temperature profile. The results show that, except for high inelasticities, the effect of inelasticity on the profiles is to slightly decrease the quantitative deviations  from the NS results.
\end{abstract}

 \maketitle
\section{Introduction}
As is well known, the Poiseuille flow is a typical example of  fluid dynamics  described in many textbooks.\cite{textbooks} In its classical formulation, the Poiseuille problem consists of finding the flow velocity and temperature profiles of a Newtonian fluid  enclosed in a slab or in a pipe  and subject to  a longitudinal pressure gradient.  Essentially the same effect is generated when the longitudinal pressure difference is replaced by a uniform gravitational force  $m \mathbf{g}$  directed longitudinally. This latter mechanism for driving the Poiseuille flow does not produce longitudinal gradients and so has proven to be  more convenient than the former in computer simulations as well as from  the theoretical point of 
view, especially to assess  the reliability of the continuum 
description.\cite{KMZ87,AS92,ELM94,TS94,TTE97,MBG97,TSS98,RC98,UG99,HM99,CR01,TS01,ATN02,ZGA02,STS03}

Kinetic theory analyses of the gravity-driven Poiseuille flow based on an expansion in powers of the gravity strength 
$g$,\cite{TS94,TSS98,TS01,STS03}  on Grad's moment method,\cite{RC98,HM99} or on an expansion in powers of the Knudsen number,\cite{ATN02} show interesting non-Newtonian effects. In particular, to second order in $g$ the temperature profile includes  a positive quadratic term,  in addition to the  negative quartic term predicted by the Navier--Stokes (NS) description. As a consequence of this new term, 
the temperature
profile exhibits a \textit{bimodal} shape with a local minimum at the middle of
the channel surrounded by two symmetric maxima at a distance of a
few mean free paths. In contrast, the NS hydrodynamic equations
predict a temperature profile with a (flat) maximum at the middle.
The Fourier law is dramatically violated since in the slab
enclosed by the two maxima the transverse component of the heat flux is
parallel (rather than anti-parallel) to the thermal gradient.
This correction to the NS temperature profile  is not captured by the next hydrodynamic description, i.e., by the Burnett equations.\cite{TSS98,UG99}
The kinetic theory  prediction of a bimodal temperature profile  has been
confirmed by numerical Monte Carlo 
simulations of the Boltzmann equation\cite{MBG97,UG99,ZGA02} and by molecular
dynamics simulations.\cite{RC98,CR01}
 On the other hand, when the Poiseuille flow is driven by a longitudinal pressure gradient instead of an external force, the NS description is in good agreement with Monte Carlo simulations of the Boltzmann equation.\cite{ZGA02}

Notwithstanding its theoretical and academic interest, the Poiseuille flow induced by gravity is of little practical interest for conventional gases under terrestrial conditions.  At a microscopic level, the relevant dimensionless parameter measuring the strength of gravity is $g\lambda/v_{\text{th}}^2$, where $\lambda$ is the mean free path and $v_{\text{th}}$ is a typical molecular speed (or thermal velocity).
The parameter $g\lambda/v_{\text{th}}^2$ measures the effect of gravity on a molecule between two successive collisions.  For instance, in the case of argon at room pressure and temperature, one has $\lambda\sim 700~\text{\AA}$ and $v_{\text{th}}\sim 400~\text{m/s}$,\cite{HCB64} so that $g\lambda/v_{\text{th}}^2\sim 5\times 10^{-12}$. 

The negligible effect of gravity on molecular gases is a consequence of their small mean free paths and large thermal velocities over mesoscopic or macroscopic scales. However, this is not necessarily so when dealing with a ``granular'' gas,\cite{C90,JN96,K99,G99,D00,BP04} i.e., a collection of a large number of discrete solid particles (or grains) in a fluidized state such that each particle moves freely and independently of the rest, except for the occurrence of inelastic binary collisions. Depending on the material properties of the grains, the solid fraction, and the state of agitation, the parameter $g\lambda/v_{\text{th}}^2$ can take values within  a wide spectrum. Let us take three representative examples. In Ref.\ \onlinecite{BK01}, the statistical properties of stainless-steel spheres of diameter $\sigma=3.175~\text{mm}$ rolling on an inclined surface and driven by an oscillating wall were experimentally studied. Typical values of the mean free path and the 
thermal velocity were $\lambda\sim 1~\text{cm}$ and $v_{\text{th}}\sim 1~\text{cm/s}$, which leads to $g\lambda/v_{\text{th}}^2\sim 10^3$.
Experiments on glass beads of diameter $\sigma=4~\text{mm}$ driven by a vertically oscillating boundary were reported in Ref.\ \onlinecite{LCDKG99}. In those experiments, $\lambda\sim\sigma$ and $v_{\text{th}}\sim  20~\text{cm/s}$, so that $g\lambda/v_{\text{th}}^2\sim 1$.
As a final example, we consider the experiments carried out in a flying rocket on bronze spheres of diameter 
$\sigma=0.3\text{--}0.4~\text{mm}$ excited by vibrations.\cite{FWEFCGB99} {}From the experimental data corresponding to the most dilute cell one can estimate $\lambda=1~\text{mm}$ and $v_{\text{th}}=5~\text{m/s}$; under terrestrial conditions 
($g=9.8~\text{m/s$^2$}$), this implies $g\lambda/v_{\text{th}}^2\sim 10^{-3}$.

In this paper we address the granular Poiseuille flow generated by gravity under the assumption that $g\lambda/v_{\text{th}}^2$ is (i) large enough as to produce noticeable gradients of density, flow velocity, and granular temperature, but (ii) small enough as to allow for a perturbative treatment; roughly speaking, this corresponds to $10^{-3}\lesssim g\lambda/v_{\text{th}}^2\lesssim 10^{-1}$.
Since kinetic energy is continuously being dissipated by inelastic collisions, we assume that the gas is externally excited by a ``heating'' mechanism. This guarantees that the gas is in a (uniform) steady state even in the absence of gravity.
As the simplest way of mimicking energy input through boundary vibrations, we consider the widely used stochastic force with white noise properties.
This means that every particle receives uncorrelated random kicks. Besides, the relative magnitude of the kicks scales with the square root of the local collision rate.

Our main goal is to derive the profiles of the hydrodynamic variables and their fluxes in the bulk region, and assess to what extent they are influenced by the degree of inelasticity.
In principle, an adequate framework to undertake this task is provided by the Boltzmann equation for inelastic hard spheres. However, its mathematical intricacy prevents one from deriving practical results, even in the elastic case, unless Grad's method with a high number of moments\cite{HM99} or the direct simulation Monte Carlo method\cite{MBG97,ZGA02} are employed. 
In order to get explicit expressions with a moderate calculation effort, we replace the Boltzmann inelastic collision operator by a much more tractable kinetic model recently proposed\cite{BDS99} as an extension to granular gases of the celebrated Bhatnagar--Gross--Krook (BGK) model for conventional gases.\cite{C88} The resulting kinetic equation is solved through second order in $g$ and the associated profiles of the hydrodynamic fields and their fluxes are derived. The results show that the same type of non-Newtonian properties that appear in the elastic case are present for granular gases as well. On the other hand, for small and moderate inelasticities, we observe that those effects tend to be slightly inhibited as the inelasticity increases.

The organization of the paper is as follows. Section \ref{sec2} is devoted to the description of the flow under study and its solution in an NS hydrodynamic description. The kinetic theory description  is presented in Sec.\ \ref{sec3}, where a perturbation expansion in powers of gravity is carried out.
The results are summarized and discussed in Sec.\ \ref{sec4}. Finally, the main conclusions of the paper are briefly
presented in Sec.\ \ref{sec5}.

\section{Statement of the problem\label{sec2}}
\subsection{Inelastic hard spheres}
Let us consider a granular gas composed of smooth \textit{inelastic} hard spheres of diameter $\sigma$, mass $m$, and coefficient of normal restitution $\alpha$. In the dilute regime, the one-particle velocity distribution function $f(\mathbf{r},\mathbf{v};t)$ obeys the (inelastic) Boltzmann equation \cite{GS95,BDS97}
\begin{equation}
\left( \partial _{t}+{\bf v\cdot \nabla }+{\mathbf{g}}\cdot\frac{\partial}{\partial \mathbf{v}} +\mathcal{F}\right)f=J[f,f],  
\label{2.1}
\end{equation}
where $\mathbf{g}$ is the acceleration due to an external force, $\mathcal{F}$ is the operator representing the action of a given heating mechanism to compensate for the collisional energy loss, and  $J[f,f]$ is the Boltzmann collision operator. Its expression is 
\beq
J[f,f]=\sigma ^{2}\int d{\bf v}_{1}\int d\widehat{\bm{\sigma}}\,\Theta (
{\bf v}_{01}\cdot \widehat{\bm{\sigma}})({\bf v}_{01}\cdot \widehat{\bm{\sigma}})
\left[ \alpha ^{-2}f({\bf v}'')f({\bf v}
_{1}'')-f({\bf v})f({\bf v}_{1})\right] ,
\label{2.2}
\eeq
where the explicit dependence of $f$ on ${\bf r}$ and $t$ has been omitted.
In Eq.\ (\ref{2.2}), 
$\Theta $ is the Heaviside step function, $\widehat{\bm{\sigma}}$ is a unit vector directed along the centers of the two colliding spheres at contact,
$\mathbf{v}_{01}=\mathbf{v}-\mathbf{v}_1$ is the relative velocity, and the pre-collisional or
restituting velocities ${\bf v}''$ and ${\bf v}_1''$ are
given by 
\begin{equation}
{\bf v}''={\bf v}-\frac{1+\alpha }{2\alpha}({\bf v}_{01}\cdot 
\widehat{\bm{\sigma}})\widehat{\bm{\sigma}},\quad {\bf v}_{1}''={\bf v}_{1}+\frac{1+\alpha }{2\alpha}({\bf v}_{01}\cdot \widehat{\bm{\sigma}
})\widehat{\bm{\sigma}}.  \label{2.3}
\end{equation}

The first few moments of the distribution function define the number density $n$, the flow velocity $\mathbf{u}$, and the \textit{granular} temperature $T$:
\begin{equation}
\left(\begin{array}{c}
n(\mathbf{r},t)\\
n(\mathbf{r},t){\bf u}(\mathbf{r},t)\\
n(\mathbf{r},t)T(\mathbf{r},t)
\end{array}\right)
=\int d{\bf v}
\left(\begin{array}{c}
1\\
{\bf v}\\
\frac{m}{3}V^2
\end{array}\right)f(\mathbf{r},\mathbf{v};t),
\label{2.4}
\eeq
where ${\bf V=v-u}$ is the velocity relative to the local flow. 
The macroscopic balance equations for the local densities of mass, momentum, 
and energy follow directly from Eq.~(\ref{2.1}) by taking velocity moments:
\beq
D_t n+n\nabla\cdot \mathbf{u}=0,
\label{b7}
\eeq
\beq
D_t\mathbf{u}+\frac{1}{mn}\nabla\cdot\mathsf{P}=\mathbf{g},
\label{b8}
\eeq
\beq
D_tT+\frac{2}{3n}\left(\nabla\cdot\mathbf{q}+\mathsf{P}:\nabla 
\mathbf{u}\right)=-(\zeta-\gamma)T.
\label{b9}
\eeq
In these equations, $D_t\equiv\partial_t+\mathbf{u}\cdot\nabla$ is the 
material time derivative,
\beq
\mathsf{P}(\mathbf{r},t)=m\int d\mathbf{v}\, \mathbf{V}\mathbf{V}f(\mathbf{r},\mathbf{v};t)
\label{b10}
\eeq
is the pressure or stress tensor,
\beq
\mathbf{q}(\mathbf{r},t)=\frac{m}{2}\int d\mathbf{v}\, 
V^2 \mathbf{V}f(\mathbf{r},\mathbf{v};t)
\label{b11}
\eeq
is the heat flux, 
\beq
\zeta(\mathbf{r},t)=
-\frac{m}{3 n(\mathbf{r},t)T(\mathbf{r},t)}
\int  d\mathbf{v}\, {V}^2J[f,f]
\label{b12}
\eeq
is the \textit{cooling} rate associated with the inelasticity of collisions, and
\beq
\gamma(\mathbf{r},t)=-\frac{m}{3 n(\mathbf{r},t)T(\mathbf{r},t)}
\int  d\mathbf{v}\, {V}^2\mathcal{F}f({\bf r},{\bf v};t)
\label{b2}
\eeq
is the \textit{heating} rate associated with the external driving $\mathcal{F}$.
Upon writing Eqs.\ (\ref{b7}) and (\ref{b8}) it has been assumed that $\mathcal{F}$ preserves the local 
number and momentum densities, i.e.,
\beq
\int d\mathbf{v}\, \mathcal{F}f({\bf r},{\bf v};t)
=\int d\mathbf{v}\, \mathbf{v}\mathcal{F}f({\bf r},{\bf v};t)=0.
\label{b1}
\eeq

Equation (\ref{b12}) shows that the cooling rate is a complicated nonlinear functional of $f$. 
By dimensional analysis, $\zeta\propto n T^{1/2}$, but the proportionality constant is an unknown function of $\alpha$.
A reasonable estimate of $\zeta$ can be obtained by replacing in Eq.\ (\ref{b12}) the actual velocity distribution function $f$ by its local Maxwellian approximation 
\begin{equation}
f_{\ell }(\mathbf{r},{\bf v};t)=n(\mathbf{r},t)\left[ \frac{m}{2\pi T(\mathbf{r},t)}
\right] ^{3/2}\exp \left[ -\frac{m\left( {\bf v}-{\bf u}(\mathbf{r},t)\right)
^{2}}{2T(\mathbf{r},t)}\right] .  
\label{23}
\end{equation}
The result is\cite{BDS99,BDKS98}
\begin{equation}
\zeta_{\ell}(\mathbf{r},t)=\nu(\mathbf{r},t) \frac{5}{12}
(1-\alpha ^{2}), \label{n3.14}
\end{equation}
where
\begin{equation}
\nu=
\frac{16}{5}n\sigma ^{2}\left( \frac{\pi T}{m}
\right) ^{1/2} \label{n3.14b}
\end{equation}
is an effective collision frequency, independent of the coefficient of restitution $\alpha$.

\subsection{Gravity-driven Poiseuille flow}
Now we assume that the granular gas is
enclosed between two 
 infinite parallel
plates normal to the $y$-axis. 
A constant external force per unit mass (e.g., gravity) ${\bf g}=-g
\widehat{\bf z}$ is applied
along a direction $\widehat{\bf z}$ parallel to the plates.
The geometry of the problem is sketched in Fig.\ \ref{sketch}.
\begin{figure}[t]
 \includegraphics[width=.60 \columnwidth]{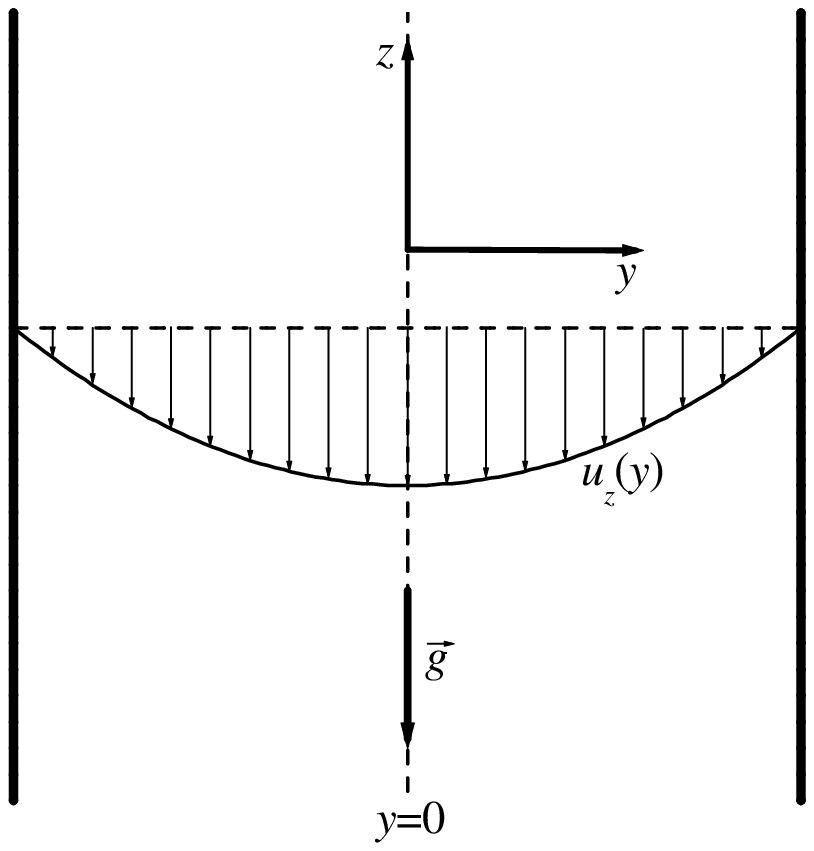}
\caption{Sketch of the planar Poiseuille flow induced by a gravitational force.
\label{sketch}}
\end{figure}

As done in laboratory experiments (and in computer simulations), we will assume that energy is externally injected into the system to compensate for the collisional
cooling, so that a steady state is achieved even if the gravity field is formally switched off.
In real experiments, \cite{BK01,LCDKG99,FWEFCGB99} this is usually achieved by means of boundary vibrations of small amplitude $A\sim \sigma$ and high frequency $\omega/2\pi\sim 10$--$100~\text{Hz}$, so that the maximum accelaration $\Gamma=A\omega^2$ is usually several times larger than the acceleration due to gravity on Earth.
However,  this type of realistic heating through the boundaries is difficult to deal with at a theoretical level due to  unavoidable boundary effects.
These difficulties are overcome by assuming a \textit{bulk} heating mechanism acting on all the particles simultaneously. 
The most commonly used type of bulk driving for inelastic particles 
consists of a stochastic force in the 
form of Gaussian white 
noise 
\cite{WM96,vNE98,SBCM98,vNETP99,BSSS99,MS00,BNK00,CCG00}. 
More precisely,  each particle $i$ is subject to
the action of a stochastic force ${\bf F}_i(t)$ that has the properties
\begin{equation}
\langle
{\bf F}_i(t)\rangle ={\bf 0},\quad
\langle
{\bf F}_i(t)
{\bf F}_j(t')\rangle
={\sf I}m^2\xi^2  \delta_{ij}\delta(t-t'),
\label{1}
\end{equation}
where ${\sf I}$ is the $3\times 3$ unit matrix and $\xi^2$ represents the strenght of the correlation.
According to this white noise driving, during a small time step $\delta t$ each particle $i$ receives an independent ``kick'' such that its velocity is incremented by a random value $\delta\mathbf{v}_i$ with the statistical properties\cite{MS00}
\beq
\langle
\delta{\bf v}_i\rangle ={\bf 0},\quad
\langle
\delta{\bf v}_i
\delta{\bf v}_j\rangle
={\sf I}\xi^2 \delta t \delta_{ij}.
\label{2.5}
\end{equation}
Therefore, $|\delta\mathbf{v}|\sim \xi\sqrt{\delta t}$.
The associated operator $\mathcal{F}$ appearing in the Boltzmann equation (\ref{2.1})  is\cite{vNE98}
\beq
\mathcal{F}=
-\frac{\xi^2}{2} 
\left(\frac{\partial}{\partial {\bf v}}\right)^2.
\label{b5}
\eeq
Thus $\xi^2/2$ plays the role of a diffusion coefficient in velocity space. 
The operator (\ref{b5}) verifies the properties (\ref{b1}), while insertion into Eq.\ (\ref{b2}) shows that the heating rate is
\beq
\gamma=\frac{m\xi^2}{T}.
\label{2.6}
\eeq

It still remains to define the spatial dependence of $\gamma$. By simplicity, we assume that the white noise driving compensates {\em
locally\/} for  the collisional energy loss. This means that $\gamma=\zeta$ or, equivalently, $\xi=\sqrt{\zeta T/m}$ at any point. This choice  can be justified by the following argument. Since, as seen above, 
$|\delta\mathbf{v}|\sim \xi\sqrt{\delta t}$, the choice $\gamma=\zeta$ implies that
\beq
\frac{|\delta\mathbf{v}|}{v_{\text{th}}}\sim \sqrt{\nu\delta t(1-\alpha^2)}, 
\label{2.7}
\eeq
where use has been made of Eq.\ (\ref{n3.14}) and of $v_{\text{th}}\sim \sqrt{T/m}$. 
Equation (\ref{2.7}) means that the \textit{relative} random increment of velocity at a given point scales as the square root of the average collision number at that point. When heating the gas  through the boundaries, the energy input is propagated to the whole system by means of collisions. Since the white noise driving intends to mimic that effect, it is quite natural that the relative magnitude of the kicks is larger in those regions where the collisions are more frequent.

By considering the above white noise excitation mechanism, 
a steady state can be expected in which
the physical
quantities depend on  the coordinate $y$ only and the flow velocity is parallel to the $z$ axis, ${\bf u}=u_z(y) \widehat{\bf z}$. In that case, the Boltzmann equation (\ref{2.1}) becomes
\beq
\left(-\frac{\zeta T}{2m}\frac{\partial^2}{\partial {\bf
v}^2}-g\frac{\partial}{\partial v_z}+v_y\frac{\partial }{\partial y}\right) f=
J[f,f].
\label{21}
\eeq
Similarly, the balance equations for momentum and energy, Eqs.\ (\ref{b8}) and (\ref{b9}), reduce to
\beq
\frac{\partial P_{yy}}{\partial y}=0,
\label{3}
\eeq
\beq
\frac{\partial P_{yz}}{\partial y}=-\rho  g,
\label{4}
\eeq
\beq
P_{yz}\frac{\partial u_z}{\partial y}+
\frac{\partial q_{y}}{\partial y}=0,
\label{5}
\eeq
where $\rho=mn$ is the mass density.
Note that the inelasticity does not appear explicitly in the
 balance equations (\ref{3})--(\ref{5}). 

\subsection{Navier--Stokes description}
In the Newtonian description the fluxes are related to the 
hydrodynamic gradients by the
 Navier--Stokes (NS) constitutive equations.\cite{BDKS98,GD99,GM01} In the geometry
 of the Poiseuille problem they read
 \beq
P_{xx}=P_{yy}=P_{zz}=p,
\label{6}
\eeq
\beq
P_{yz}=-\eta\frac{\partial u_{z}}{\partial y},
\label{7}
\eeq
\beq
q_{y}=-\kappa\frac{\partial T}{\partial y}-\mu \frac{\partial n}{\partial y},
\label{8}
\eeq
\beq
q_z=0,
\label{9}
\eeq
where $p=nT=\frac{1}{3}\mbox{Tr}\, {\sf P}$ is the hydrostatic pressure,
 $\eta$ is the the shear viscosity,  $\kappa$ is the
thermal conductivity, and $\mu$ is a transport coefficient with no analog for
elastic fluids.
These transport coefficients can be explicitly derived from the Boltzmann equation (\ref{2.1}) by application of the Chapman--Enskog method in the first Sonine approximation.  In the case of the white noise heating (\ref{b5}) with $\xi=\sqrt{\zeta T/m}$ their expressions are\cite{GM01} 
\beq
\eta=\frac{p}{ \nu_\eta}, \quad \kappa=\frac{5p}{2m\nu_\kappa }\left(1+2 k\right),
\quad \mu=\frac{5T^2}{2m \nu_\kappa }{k},
\label{B1}
\eeq
where
\beq
\nu_\eta=\frac{\nu}{4}(1+\alpha)\left(3-\alpha\right)\left(1 
-\frac{1}{32}k\right),
\label{B6}
\eeq
\beq
\nu_\kappa=\frac{\nu}{3}(1+\alpha)\left(\frac{49-33\alpha}{16}+ 
\frac{19-3\alpha}{512}k\right).
\label{B7}
\eeq
In the above equations, $\nu$ is the effective collision frequency defined by Eq.\ (\ref{n3.14b}) and $k$ is the kurtosis of the homogeneous heated state. Its expression is well approximated by\cite{vNE98}
\beq
k=\frac{16(1-\alpha)(1-2\alpha^2)}{241-177\alpha+30\alpha^2(1-\alpha)}.
\label{2.8}
\eeq
The kurtosis $k$ is rather small for all $\alpha$. In particular, $|k|<0.013$ for $0.6\leq\alpha\leq 1$. Therefore, one can neglect $k$ in (\ref{B1})--(\ref{B7}) to get
\beq
\eta\simeq \frac{p}{ \nu}\frac{4}{(1+\alpha)\left(3-\alpha\right)}, \quad \kappa\simeq\frac{5p}{2m\nu}\frac{48}{(1+\alpha)\left(49-33\alpha\right)},
\quad \mu\simeq 0.
\label{B1new}
\eeq
In the interval $0.6\leq\alpha\leq 1$, the expressions (\ref{B1new}) for $\eta$ and $\kappa$ deviate from those of (\ref{B1}) less than 0.04\% and 3\%, respectively. Besides, the ratio $n\mu/T\kappa$ is smaller than 0.013, so that $\mu$ can be neglected. Note that the negligible role played by $\mu$ does not hold in the homogeneous cooling state.\cite{BDKS98,GD99}
It is worth pointing out that, while the shear viscosity monotonically increases with inelasticity, the thermal conductivity starts decreasing with increasing inelasticity, reaches a minimum value around $\alpha=0.4$, and then slightly increases for
$\alpha\gtrsim 0.4$. This non-monotonic behavior of $\kappa$ in the heated state contrasts with the one found
in the free cooling case.\cite{BDKS98,GM01,S03}

Combining Eqs.\ (\ref{3})--(\ref{8}), we get
\beq
\frac{\partial p}{\partial y}=0,
\label{10}
\eeq
\beq
\frac{\partial}{\partial y}\eta\frac{\partial u_z}{\partial y}=\rho g,
\label{11}
\eeq
\beq
\frac{\partial}{\partial y}\kappa'\frac{\partial T}{\partial y}=-
\eta\left(\frac{\partial u_z}{\partial y}\right)^2,
\label{12}
\eeq
where $\kappa'=\kappa-n\mu/T\simeq \kappa$.
Equation (\ref{11}) gives a parabolic-like velocity profile, that is
characteristic of the Poiseuille flow.
The temperature profile has, according to Eq.\ (\ref{12}), a quartic-like
shape.
Strictly speaking, these NS profiles are more complicated than just
polynomials
due to the temperature dependence of the transport coefficients.
Since the hydrodynamic profiles must be symmetric with respect to the middle plane
$y=0$, their odd derivatives must vanish at $y=0$. Thus, from Eqs.\
(\ref{11}) and (\ref{12}) we have
\beq
\left.\frac{\partial^2 u_z}{\partial y^2}\right|_{y=0}=
\frac{\rho_0g}{\eta_0}, \quad
\left.\frac{\partial^2 T}{\partial y^2}\right|_{y=0}=0,\quad
\left.\frac{\partial^4 T}{\partial y^4}\right|_{y=0}=-2
\frac{\rho_0^2g^2}{\eta_0\kappa_0'},
\label{15}
\eeq
where  the subscript $0$  denotes quantities evaluated at $y=0$.
According to  Eq.\ (\ref{15}), the NS equations
predict that   the temperature has
a maximum at the middle layer $y=0$. 
As we will see in Sec.\ \ref{sec3}, the kinetic theory description shows
that the temperature actually exhibits a local {\em minimum\/} at $y=0$,
since $\left.\partial^2T/\partial y^2\right|_{y=0}$  is a positive quantity
(of order $g^2$).

The closed set of nonlinear equations (\ref{10})--(\ref{12}) cannot be solved
analytically for arbitrary $g$ due to the spatial dependence of the transport coefficients. 
On the other hand, if the acceleration of gravity is sufficiently small at the microscopic scale, we can  expand in powers of $g$ and keep the first few terms only.
 To second order, the NS hydrodynamic profiles near the layer $y=0$ are
\beq
u_z(y)=u_0+\frac{\rho_0
g}{2\eta_0}{y}^2+
{\cal O}(g^3),
\label{16}
\eeq
\beq
T(y)=T_0-\frac{\rho_0^2 g^2}{12\eta_0\kappa_0'}{y}^4
+{\cal O}(g^4).
\label{17}
\eeq
The space variable $y$ can be eliminated between Eqs.\
(\ref{16}) and (\ref{17}) to obtain the following
nonequilibrium ``equation of state'':
\beq
T=T_0-\frac{\eta_0}{3\kappa_0'}(u_0-u_z)^2+{\cal O}(g^4).
\label{20}
\eeq
The NS profiles for the fluxes are
\beq
P_{yz}(y)=-\rho_0 g{y}+{\cal O}(g^3),
\label{18}
\eeq
\beq
q_{y}(y)=\frac{\rho_0^2 g^2}{3\eta_0}{y}^3+{\cal O}(g^4).
\label{19}
\eeq
\section{Kinetic theory description\label{sec3}}
\subsection{A kinetic model}
In this Section we will see that most of the NS predictions discussed in the preceding Subsection do not hold true, even to first order in $g$, when the problem is attacked from a more detailed kinetic point of view.
In principle, the task consists of solving the Boltzmann equation (\ref{21}) through order $g^2$ in a region near the central layer $y=0$.

Given the mathematical complexity of the Boltzmann
collision operator (\ref{2.2}), especially in the case of inelastic collisions,  we simplify the analysis by  replacing $J[f,f]$ by a BGK-like kinetic model:\cite{BDS99,SA04}
\beq 
J[f,f]\to -\beta(\alpha)\nu (f-f_{\ell })+\frac{\zeta_{\ell }}{2}
\frac{\partial}{\partial{\bf v}}\cdot \left[ \left( {\bf v-u}\right) f\right] ,
\label{22}
\end{equation}
where $\nu$ is the collision frequency (\ref{n3.14b}), $f_{\ell }$ is the local Maxwellian distribution (\ref{23}), and $\zeta_{\ell}$ is the associated cooling rate (\ref{n3.14}). In addition, $\beta(\alpha)$ is a dimensionless function of the coefficient of restitution that can be freely chosen to optimize agreement with the  Boltzmann description.
Equation (\ref{22}) differs from the original formulation of the model kinetic equation\cite{BDS99}  in that the exact (local) homogeneous cooling state of the Boltzmann equation is replaced by $f_\ell$ and the exact cooling rate (\ref{b12}) is approximated by $\zeta_\ell$. Confirmation of the quantitative agreement between the
kinetic model and the Boltzmann equation has been found for the simple shear flow\cite{BRM97,MGSB99} and the nonlinear Couette flow.\cite{TTMGSD01}

 The first term on the right-hand side of (\ref{22})
 describes collisional relaxation towards the local
Maxwellian with a collision rate $\beta\nu $, while the second term describes the
dominant collisional cooling effects. 
The necessity for this term to accurately represent the spectrum of the
Boltzmann collision operator is discussed in Ref.\ \onlinecite{BDS99}. However, 
it can be
viewed more simply as an effective ``drag'' force that produces the same
energy loss rate as that produced by the inelastic collisions. The NS transport coefficients derived from the model (\ref{22}) in the case of white noise heating are\cite{S03}
\beq
\eta=\frac{p}{\beta \nu+\zeta_\ell}, \quad \kappa=\frac{5p}{2m\left(\beta\nu+\frac{3}{2}\zeta_\ell\right)},
\quad \mu=0.
\label{3.1}
\eeq
A simple choice for the parameter $\beta$ is $\beta=\frac{1}{2}(1+\alpha)$.\cite{SA04} On the other hand,
comparison with the (approximate) Boltzmann results (\ref{B1new}) shows that the shear viscosity  is reproduced if $\beta$ takes the value
\beq
\beta=(1+\alpha)\frac{2+\alpha}{6},
\label{3.2}
\eeq
while the thermal conductivity is reproduced if
\beq
\beta=(1+\alpha)\frac{19-3\alpha}{48}.
\label{3.3}
\eeq
The discrepancy between Eqs.\ (\ref{3.2}) and (\ref{3.3}) persists in the elastic limit ($\alpha=1$) and is a well-known
limitation of the BGK model.
As will be seen in Sec.\ \ref{sec4}, one can partially circumvent this problem by expressing the final results in terms of $\eta$ and $\kappa$.

Inserting the model (\ref{22}) into Eq.\ (\ref{21}), we get the kinetic
equation
\beq
\left(-g\frac{\partial}{\partial v_z}+v_y\frac{\partial }{\partial y}\right) f=
-\beta\nu (f-f_{\ell })+\frac{\zeta_\ell}{2}
\frac{\partial}{\partial{\bf v}}\cdot \left( {\bf
V}+\frac{T}{m}\frac{\partial}{\partial{\bf v}}\right) f , 
\label{27}
\eeq
where, for consistency, we have made the approximation $\zeta\to\zeta_\ell$ in Eq.\ (\ref{21}).
In order to focus on the deviations from the local equilibrium distribution, let
us write
\beq
f=f_\ell(1+\Phi).
\label{27.2}
\eeq
Then, Eq.\ (\ref{27}) becomes
\beqa
&&(1+\Phi)\left[V_y
\widetilde{\partial}_y \log f_\ell-\left(g+V_y\frac{\partial u_z}{\partial
y}\right)\frac{\partial}{\partial V_z}\log f_\ell\right]=\nonumber\\
&&\left(g+V_y\frac{\partial u_z}{\partial
y}\right)\frac{\partial}{\partial V_z} \Phi -V_y
\widetilde{\partial}_y\Phi
-(\nu'-\zeta_\ell)\Phi
+\frac{\zeta_\ell}{2}\left(
\frac{T}{m}\frac{\partial}{\partial{\bf
V}}-{\bf V}\right)\cdot\frac{\partial}{\partial{\bf V}}\Phi,\nn
\label{27.3}
\eeqa
where the operator $\widetilde{\partial}_y\equiv \partial/\partial y+(\partial
u_z/\partial y)\partial/\partial V_z$ derives with respect to $y$ at constant ${\bf V}$
(i.e., not at constant ${\bf v}$).
Moreover, in Eq.\ (\ref{27.3}) we have introduced the modified collision frequency $\nu'\equiv\beta\nu+\zeta_\ell$. As Eq.\ (\ref{3.1}) shows, $\nu'$ is the effective collision frequency associated with the shear viscosity of the heated granular gas in the kinetic model. 

Since we are interested in the solution of Eq.\ (\ref{27.3}) in the bulk, it is convenient to take the state at the mid point
$y=0$ as a reference state and define the following dimensionless quantities:
\beq
{\bf V}^*={\bf V}/v_0,\quad y^*=y\nu_0'/v_0,\quad  f_\ell^*=f_\ell v_0^3/n_0,
\label{28}
\eeq
\beq
p^*=p/p_0, \quad {\bf u}^*={\bf u}/v_0,\quad T^*=T/T_0, \quad g^*=g/\nu_0'v_0
\label{29}
\eeq
\beq
{\nu'}^*=\nu'/\nu_0',\quad \mathsf{P}^*={\sf
P}/p_0, \quad {\bf q}^*={\bf q}/p_0 v_0,
\label{30}
\eeq
where, as in Eqs.\ (\ref{15})--(\ref{19}), the subscript 0 denotes quantities at
$y=0$. In particular, $v_0=(2T_0/m)^{1/2}$ is the thermal velocity $v_{\text{th}}$ at $y=0$. The reduced quantity $y^*$ measures distance in units of a nominal mean free path, while $g^*$ measures the strength of the gravity field on a particle moving with the thermal velocity along a distance of the order of the mean free path. The choice of $1/\nu_0'$ (which depends on $\alpha$) instead of $1/\nu_0$ (which is independent of $\alpha$) as the time unit is suggested by a larger simplicity in the calculations stemming from the kinetic model. In any case, in Section \ref{sec4} we will summarize the results in real units, so the final expressions are independent of the specific choice of reduced quantities.

In the above units, the kinetic equation (\ref{27.3}) becomes
\beqa
&&(1+\Phi)\left[V_y^*
\widetilde{\partial}_{y^*} \log f_\ell^*+\frac{2V_z^*}{T^*}\left(g^*+V_y^*\frac{\partial u_z^*}{\partial
y^*}\right)\right]=\left(g^*+V_y^*\frac{\partial u_z^*}{\partial
y^*}\right)\frac{\partial}{\partial V_z^*} \Phi
\nonumber\\
&& -V_y^*
\widetilde{\partial}_{y^*}\Phi-{\nu'}^*\left(1-\z\right)\Phi
+\z\frac{{\nu'}^*}{2}\left(
\frac{T^*}{2}\frac{\partial}{\partial{\bf
V}^*}-{\bf V}^*\right)\cdot\frac{\partial}{\partial{\bf V}^*}\Phi,\nn
\label{31}
\eeqa
where 
\begin{equation}
\widetilde{\partial}_{y^*} \log f_\ell^*=\frac{\partial \log p^*}{\partial y^*}
+\left(\frac{V^2}{T}-\frac{5}{2}\right)\frac{\partial \log T^*}{\partial y^*}.
\label{31.1}
\end{equation}
On the right-hand side of Eq.\ (\ref{31}) we have taken into account that $\zeta_\ell=\z \nu'$, where [cf.\ Eq.\ (\ref{n3.14})]
\beq
\z=\frac{\frac{5}{12}(1-\alpha^2)}{\beta(\alpha)+\frac{5}{12}(1-\alpha^2)}
\label{zeta0}
\eeq
is a pure number that only depends on the coefficient of restitution.
It gives the cooling rate at any given point in units of the modified collision frequency $\nu'$ at that same point.

Our purpose is to solve  Eq.\ (\ref{31}) to second order in $g^*$ and get the
associated hydrodynamic profiles.

\subsection{Perturbation expansion}
In this Subsection, all the quantities will be understood to
be expressed in reduced units and the asterisks will be dropped.
The expansion of $\Phi$  in powers of $g$ is
\beq
\Phi=\Phi^\one g+\Phi^\two g^2+{\cal O} (g^3),
\label{32}
\eeq
where we have taken into account that the solution of Eq.\ (\ref{27}) in the absence of gravity is $f=f_\ell$  with uniform $n$, $\mathbf{u}$, and $T$.
The  expansions for the hydrodynamic
fields have the forms
\beq
p=1+p^\two g^2+{\cal O}( g^4),
\label{33}
\eeq
\beq
u_z=u^\one g+{\cal O}( g^3),
\label{34}
\eeq
\beq
T=1+T^\two g^2+{\cal O}( g^4).
\label{35}
\eeq
Here we have taken into account that, because of the symmetry of the
problem, $p$ and $T$ are even functions of $g$, while $u_z$ is an odd
function. Also, without loss of generality, we have taken $u_0=0$, i.e., we are performing a Galilean change to a reference frame moving with the fluid at $y=0$. Since
$\nu'=pT^{-1/2}$, we have
\beq
\nu'=1+\left(p^\two-\frac{1}{2}T^\two\right) g^2+{\cal O}( g^4).
\label{36}
\eeq
Nevertheless, only $\nu'=1$ is needed in the evaluation of $\Phi^{(1)}$ and $\Phi^{(2)}$.

In order to solve Eq.\ (\ref{31}) at each order, we will need to use the
consistency conditions
\beq
\int d{\bf V}\, f_\ell \Phi=0,
\label{37}
\eeq
\beq
\int d{\bf V}\, V_y f_\ell \Phi=0,
\label{38.1}
\eeq
\beq
\int d{\bf V}\, V_z f_\ell \Phi=0,
\label{38}
\eeq
\beq
\int d{\bf V}\, V^2 f_\ell \Phi=0.
\label{39}
\eeq

\subsubsection{First order}
To first order in $g$, Eq.\ (\ref{31}) yields
\beq
\left(1-{\cal
A}\right)\Phi^\one=-\frac{2}{1-\z}V_z \left(1+V_y\frac{\partial
u^\one}{\partial
y}\right)
\equiv\phi^\one,
\label{43}
\eeq
where ${\cal A}$ is the operator
\beq
{\cal A}=\frac{\z}{2(1-\z)}
\left(
\frac{1}{2}\frac{\partial}{\partial{\bf V}}-{\bf V}\right)\cdot\frac{\partial}{\partial{\bf V}}
-\frac{1}{1-\z}V_y
\widetilde{\partial}_y .
\label{44}
\eeq
The function $\phi^\one$ has
a known velocity dependence. Its space dependence occurs through $u^\one$,
which remains unknown so far. In order to solve Eq.\ (\ref{43}), we will follow
a heuristic procedure. First, we guess that the first-order velocity profile is
parabolic:
\beq
u^\one(y)=u^\one_2 y^2.
\label{45}
\eeq
Next, we note that the formal solution to Eq.\ (\ref{43}) is $\Phi^\one
=\sum_{k=0}^\infty {\cal A}^k \phi^\one$ and that the functional structure of
${\cal A}^k \phi^\one$ remains the same for any $k$. Consequently, the solution
to Eq.\ (\ref{43}) must have such a structure, namely
\beq
\Phi^\one(y,\mathbf{V})=V_z(a_0+a_1 V_y^2+a_2 V_y y).
\label{46}
\eeq
Equations (\ref{45}) and (\ref{46}) have the same structure as the
solution of the BGK equation in the elastic case.\cite{TS94,TS01} Insertion of
Eq.\ (\ref{46}) into Eq.\ (\ref{43}) allows one to identify the coefficients
$a_0,a_1,a_2$. The result is
\beq
a_0=4\frac{2\z u^\one_2-\z-2}{4-{\z}^2}
, \quad a_1=\frac{8u^\one_2}{2+\z}, \quad a_2=-4u^\one_2.
\label{47}
\eeq
The consistency conditions (\ref{37}), (\ref{38.1}), and (\ref{39}) are verified by symmetry.
The coefficient $u^\one_2$ is determined by the condition (\ref{38}) with the
result 
\beq
u^\one_2=1.
\label{45bis}
\eeq

 Once we know $\Phi^\one$ explicitly, we can get the non-zero
components of the fluxes to first order. They are
\beq
P_{yz}^\one(y)=2\int d{\bf V}\, V_y V_z f_0\Phi^\one=-2y
\label{48}
\eeq
\beq
q_{z}^\one(y)=\int d{\bf V}\, V^2 V_z f_0\Phi^\one
=\frac{2}{2+\z},
\label{49}
\eeq
where $f_0=\pi^{-3/2}e^{-V^2}$ is $f_\ell$ at $y=0$.
\subsubsection{Second order}
We proceed in a similar way as before. The equation for $\Phi^\two$ is
\beqa
\left(1-{\cal
A}\right)\Phi^\two&=&\frac{1}{1-\z} \left(1+V_y\frac{\partial
u^\one}{\partial
y}\right)\left(\frac{\partial}{\partial V_z}-2V_z\right)\Phi^\one\nn
&&-
\frac{V_y}{1-\z}\left[\frac{\partial  p^\two}{\partial y}
+\left({V^2}-\frac{5}{2}\right)\frac{\partial T^\two}{\partial y}\right]
\equiv\phi^\two.
\label{50}
\eeqa
Now, we guess the profiles
\beq
p^\two(y)=p^\two_2 y^2,
\label{51}
\eeq
\beq
T^\two(y)=T^\two_2 y^2+T^\two_4 y^4.
\label{55}
\eeq
The structure of ${\cal A}^k \phi^\two$ suggests the trial function
\beqa
\Phi^\two(y,\mathbf{V})&=&b_0+b_1 V_y^2+b_2 V_y y+b_3 y^2+b_4 V_y^4+b_5 V_y^3 y+b_6V_y^2 y^2\nn
&& 
+b_7V_y y^3
+V_z^2\left(c_0+c_1 V_y^2+c_2 V_y y+c_3 y^2+c_4 V_y^4\right.\nn
&&\left.+c_5 V_y^3 y+c_6V_y^2 y^2
\right)
+V^2\left(d_0+d_1 V_y^2+d_2 V_y y+d_3 y^2\right.\nn
&&\left.+d_4 V_y^4+d_5 V_y^3 y+d_6V_y^2 y^2
+d_7V_y y^3\right).
\label{56}
\eeqa
Insertion into Eq.\ (\ref{50}) allows one to get the coefficients $b_i$, $c_i$, and $d_i$ in terms of
$p_2^\two$, $T_2^\two$, and $T_4^\two$. Condition (\ref{38.1}) is identically
satisfied regardless of the values of $p_2^\two$, $T_2^\two$, and $T_4^\two$, while
Eq.\ (\ref{38}) is verified by symmetry. On the other hand, conditions
(\ref{37}) and (\ref{39}) yield
\beq
p_2^\two=\frac{24}{5}, \quad
T_2^\two=\frac{4}{25}\frac{38+43\z+17{\z}^2}{(1+\z)(2+\z)}, \quad
T_4^\two=-\frac{2}{15}(2+\z).
\label{57}
\eeq
The expressions of the coefficients $b_i$, $c_i$, and $d_i$ as functions of $\alpha$ are given in 
Appendix \ref{appA}.
{}From $\Phi^\two$ we can calculate the second order contributions to the
fluxes:
\beq
P_{yy}^\two=p^\two+2\int d{\bf V}\, V_y^2 f_0\Phi^\two=-\frac{24}{25}\frac{102+87\z+13{\z}^2}{(1+\z)(2+\z)^2},
\label{58}
\eeq
\beq
P_{zz}^\two(y)=p^\two+2\int d{\bf V}\, V_z^2 f_0\Phi^\two=\frac{32}{25}\frac{82+67\z+8{\z}^2}{(1+\z)(2+\z)^2}+\frac{56}{5}y^2,
\label{59}
\eeq
\beq
q_{y}^\two(y)=\int d{\bf V}\,V^2 V_y^2 f_0\Phi^\two=\frac{4}{3}y^3.
\label{60}
\eeq
\section{Summary and discussion\label{sec4}}
\subsection{Hydrodynamic profiles}
Let us summarize here the main results obtained from the kinetic model through second order in the gravity field. The
hydrodynamic profiles are given by Eqs.\ (\ref{33})--(\ref{35}), (\ref{45}), (\ref{45bis}), (\ref{51}), (\ref{55}), and (\ref{57}). Expressed in real units, they are
\beq
p(y)=p_0\left[1+\frac{6}{5}\left(\frac{mg}{T_0}\right)^2 y^2\right]+{\cal
O}(g^4),
\label{61}
\eeq
\beq
u_z(y)=u_0+\frac{\rho_0
g}{2\eta_0}{y}^2+
{\cal O}(g^3),
\label{62}
\eeq
\beq
T(y)=T_0\left[1-\frac{\rho_0^2
g^2}{12\eta_0\kappa_0T_0}{y}^4+
\frac{1}{25}\frac{38+43\z+17{\z}^2}{(1+\z)(2+\z)}\left(\frac{mg}{T_0}\right)^2
y^2\right] 
+{\cal O}(g^4).
\label{63}
\eeq
In Eqs.\ (\ref{62}) and (\ref{63}), $\eta_0$ and $\kappa_0=\kappa'_0$ are the NS transport coefficients (evaluated at the mid point $y=0$) of the granular gas heated by the stochastic force. In the kinetic model, those transport coefficients are given by Eq.\ (\ref{3.1}). 
Note that the elimination of the collision frequencies $\nu$ or $\nu'=\beta\nu+\zeta_\ell$ in favor of the transport coefficients $\eta$ and $\kappa$ allows us to circumvent the limitation inherent to BGK-like models of not giving the correct Prandtl number. 
In that way, Eqs.\ (\ref{61})--(\ref{63}) can be expected to be close to the Boltzmann results, as happens in the elastic case.\cite{TSS98,STS03}
Therefore, in what follows we will use for $\eta$ and $\kappa$ the Boltzmann expressions (\ref{B1new}). In Eq.\ (\ref{63}) the parameter $\z$ is given by Eq.\ (\ref{zeta0}), where $\beta(\alpha)$ can be freely chosen. Here we will take the choice (\ref{3.2}), which makes the NS shear viscosity of the kinetic model agree with that of the Boltzmann equation.

Comparison of Eqs.\ (\ref{61})--(\ref{63}) with the NS predictions, Eqs.\
(\ref{10}), (\ref{16}), and (\ref{17}) shows that the latter provide an
incomplete description to second order in $g$. According to the kinetic
theory description, the pressure increases
parabolically from the mid layer rather than being uniform and the temperature
has an extra positive quadratic term that is responsible for the fact that the
temperature has a local minimum at $y=0$ rather than a maximum. This minimum
is surrounded by two symmetric maxima located at a distance ($y_{\text{max}}$) 
from $y=0$  of the order of a few mean free paths.
Analogously, the NS equation of state (\ref{20}) is corrected by an extra term,
\beq
T=T_0-\frac{\eta_0}{3\kappa_0}(u_0-u_z)^2+
\frac{1}{30}\frac{38+43\z+17{\z}^2}{(1+\z)(2+\z)}(p-p_0)+{\cal O}(g^4).
\label{64}
\eeq

\begin{figure}[t]
 \includegraphics[width=.80 \columnwidth]{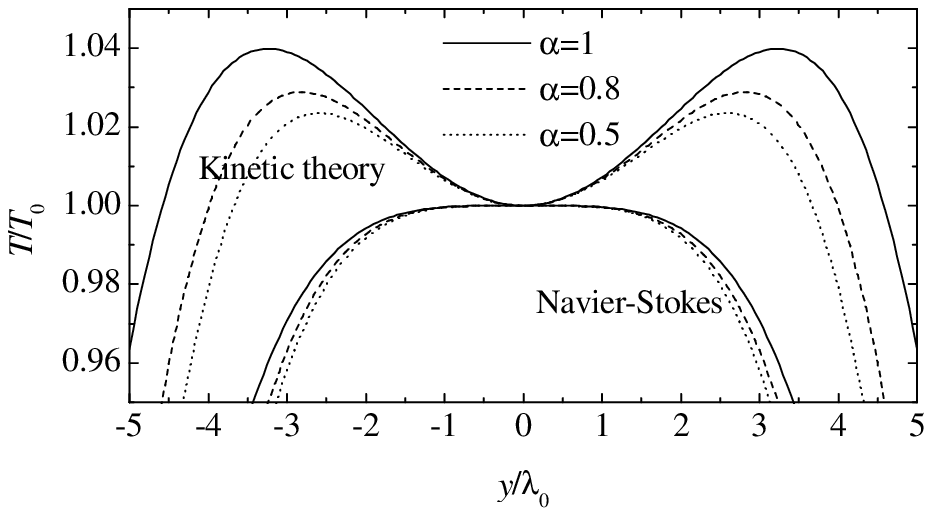}
\caption{Temperature profiles  for $g\lambda_0/v_0^2=0.05$ and $\alpha=0.5$ (dotted lines), $\alpha=0.8$ (dashed lines), and $\alpha=1$ (solid lines), as predicted by the Navier--Stokes and kinetic theory descriptions.
\label{Tprofile}}
\end{figure}
\begin{figure}[t]
 \includegraphics[width=.80 \columnwidth]{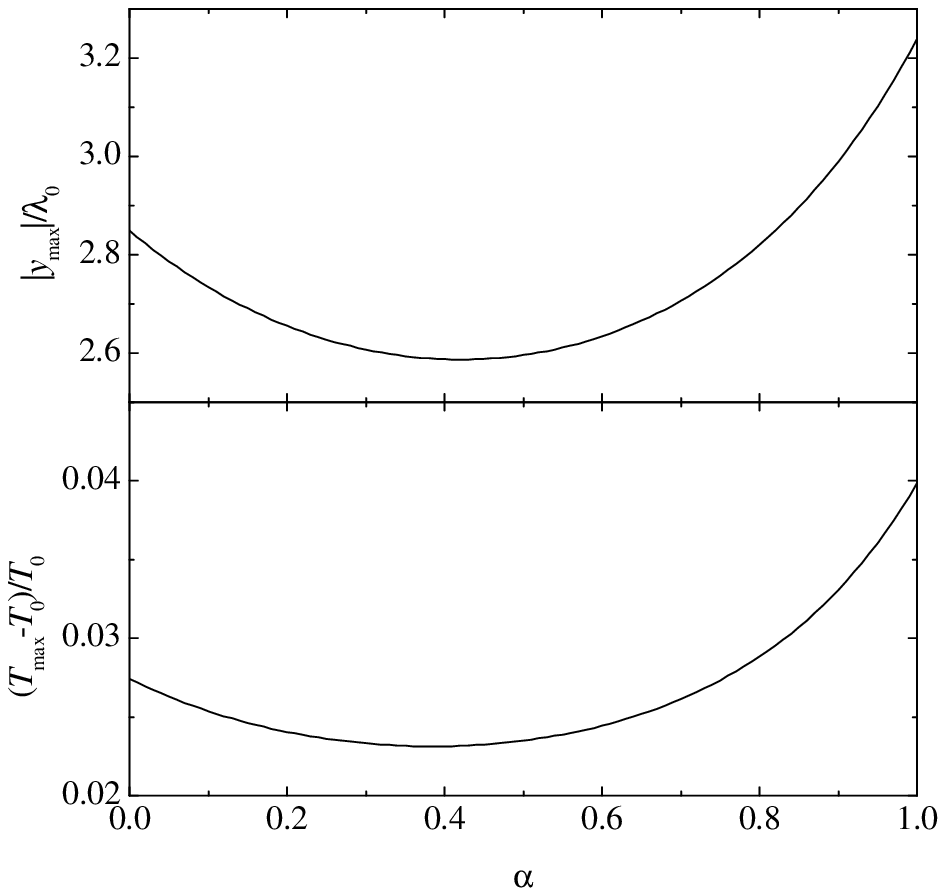}
\caption{Plot of $|y_{\text{max}}|/\lambda_0$ and $(T_{\text{max}}-T_0)/T_0$ versus $\alpha$. In the latter case we have taken $g\lambda_0/v_0^2=0.05$.
\label{ymaxTmax}}
\end{figure}

In order to analyze in detail the temperature profile (\ref{63}), let us measure the coordinate $y$ in units of the mean free path $\lambda_0=(\pi\sqrt{2}n_0\sigma^2)^{-1}=(8/5\sqrt{\pi})v_0/\nu_0$, where $v_0=\sqrt{2T_0/m}$ is the thermal velocity and $\nu_0$ is the collision frequency (\ref{n3.14b}), both at $y=0$. Thus, Eq.\ (\ref{64}) becomes
\beq
\frac{T(y)}{T_0}=1-A_4(\alpha)\left(\frac{g\lambda_0}{v_0^2}\right)^2\left(\frac{y}{\lambda_0}\right)^4+A_2(\alpha)\left(\frac{g\lambda_0}{v_0^2}\right)^2\left(\frac{y}{\lambda_0}\right)^2+\mathcal{O}(g^4),
\label{4.1}
\eeq
where
\beq
A_4(\alpha)=\frac{4}{1125\pi}(1+\alpha)^2(3-\alpha)(49-33\alpha),\quad
A_2=\frac{4}{25}\frac{2719-2741\alpha+706\alpha^2}{(7-4\alpha)(23-11\alpha)}.
\label{4.2}
\eeq
The coefficient $A_2(\alpha)$ monotonically decreases with increasing inelasticity, while $A_4(\alpha)$ has a maximum at $\alpha\simeq 0.46$, essentially due to the non-monotonic behavior of the thermal conductivity.

The location $y_{\text{max}}$ of the two symmetric maxima is
\beq
y_{\text{max}}=\pm \lambda_0\sqrt{\frac{A_2(\alpha)}{2A_4(\alpha)}}.
\label{4.3}
\eeq
Note that, in the regime $g\lambda_0/v_0^2\ll 1$, $y_{\text{max}}$ is independent of the precise value of $g$. The relative value of the maximum temperature is
\beq
\frac{T_{\text{max}}-T_0}{T_0}=\frac{A_2^2(\alpha)}{4A_4(\alpha)}\left(\frac{g\lambda_0}{v_0^2}\right)^2+\mathcal{O}(g^4).
\label{4.4}
\eeq
Of course,  if we formally make  $A_2(\alpha)=0$ in Eq.\ (\ref{4.1}), the NS temperature profile (\ref{17}) is recovered.

As an illustration of the corrections over the NS description provided by the
kinetic model, let us consider a value $g\lambda_0/v_0^2=0.05$. In the case of terrestrial gravity, the above value corresponds, for instance, to $\lambda_0\sim 5~\text{mm}$ and $v_0\sim 1~\text{m/s}$. Although terms of order
higher than $g^2$ in (\ref{4.1}) might not be negligible for this particular value of $g\lambda_0/v_0^2$, the qualitative features  are expected to remain correct.
Figure \ref{Tprofile} shows the temperature profiles for a granular gas with $\alpha=0.5$ and $\alpha=0.8$, as well as for a gas of elastic particles ($\alpha=1$), as predicted by the NS and kinetic theory descriptions.
We observe that strong deviations from the NS profiles are apparent, both for  elastic and  inelastic systems. Focusing now on the profiles predicted by the kinetic model, we se that, as the inelasticity increases, the locations of the two maxima
shift towards the center of the slab and the value of the maximum temperature
 decreases. 
This behavior, however, is reversed if $\alpha\lesssim 0.4$. The $\alpha$-dependence of $|y_{\text{max}}|/\lambda_0$ and $(T_{\text{max}}-T_0)/T_0$ is displayed in Fig.\ \ref{ymaxTmax}. The non-monotonic behaviors of $y_{\text{max}}$ and $T_{\text{max}}$ are consequences of that of $A_4(\alpha)$.

\subsection{Fluxes}
The profiles for the elements of the pressure tensor and the components of the heat flux through second order are given by Eqs.\ (\ref{48}), (\ref{49}), and (\ref{58})--(\ref{60}). Expressed in real units, they are 
\beq
P_{yz}(y)=-\rho_0 g{y}+{\cal O}(g^3),
\label{65}
\eeq
\beq
P_{yy}=p_0\left[1-\frac{12}{25}\frac{102+87\z+13{\z}^2}{(1+\z)(2+\z)^2}
\frac{\rho_0\eta_0^2g^2}{p_0^3}\right]+{\cal O}(g^4),
\label{66}
\eeq
\beq
P_{zz}(y)=p_0\left[1+\frac{16}{25}\frac{82+67\z+8{\z}^2}{(1+\z)(2+\z)^2}
\frac{\rho_0\eta_0^2g^2}{p_0^3}+
\frac{14}{5}\left(\frac{mg}{T_0}\right)^2 y^2\right]+{\cal O}(g^4),
\label{67}
\eeq
\beq
q_{y}(y)=\frac{\rho_0^2 g^2}{3\eta_0}{y}^3+{\cal O}(g^4),
\label{68}
\eeq
\beq
q_z=\frac{2}{5}m\kappa_0 g+{\cal O}(g^3).
\label{69}
\eeq
The $xx$-element of the pressure tensor is $P_{xx}=3p-P_{yy}-P_{zz}$. Note that $P_{yy}$ is uniform, in agreement with the exact balance equation (\ref{3}). Likewise, it is easy to check that Eqs.\ (\ref{62}), (\ref{65}), and (\ref{68}) are consistent with the energy balance equation (\ref{5}). Moreover, since the density profile is known through second order [cf.\ Eqs.\ (\ref{61}) and (\ref{63})], Eq.\ (\ref{4}) can be used to get $P_{yz}$ through third order. 

The shear stress $P_{yz}$ agrees to second order in $g$ with Newton's  viscosity law (\ref{7}). However, the component $q_y$ of the heat flux  parallel to the thermal gradient does not obey Fourier's law (\ref{8}) (note that $\mu\approx 0$ in the heated state). In fact, from Eqs.\ (\ref{63}) and (\ref{68}) one can write an 
\beq
q_y=-\kappa\frac{\partial }{\partial y}\left(T+\frac{y_{\text{max}}^2}{6}\nabla^2 T\right) +\mathcal{O}(g^4),
\label{4.5}
\eeq
which shows that one needs to incorporate super-Burnett contributions to account for the relationship between the heat flux and the thermal gradients. The extra term on the right-hand side of Eq.\ (\ref{4.5}) is responsible for the counter-intuitive fact of $q_y$ having the same sign as $\partial T/\partial y$ in the region $0\leq |y|<|y_{\text{max}}|$, i.e., the temperature increases as one moves away from the mid layer $y=0$ and yet the heat flows outward from the colder to the hotter layers. A steady state is still possible because the energy deficit is compensated for  by the viscous heating. 
An additional departure from Fourier's law is related to the existence of a component $q_z$ of the heat flux normal to the thermal gradient, an effect that is already of first order in $g$ and is related to a Burnett contribution associated with $\nabla^2 u_z$.\cite{TSS98}

\begin{figure}[t]
 \includegraphics[width=.80 \columnwidth]{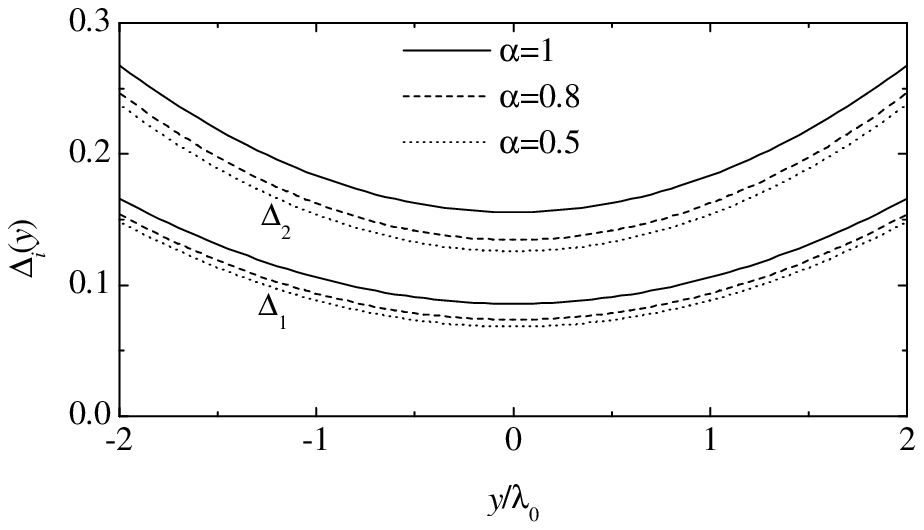}
\caption{Profiles of the normal stress differences $\Delta_{1}=(P_{zz}-P_{xx})/{p}$ and $\Delta_{2}=(P_{zz}-P_{yy})/{p}$ for $g\lambda_0/v_0^2=0.05$ and $\alpha=0.5$ (dotted lines), $\alpha=0.8$ (dashed lines), and $\alpha=1$ (solid lines).
Note that $\Delta_1=\Delta_2=0$ in the NS description.
\label{Delta12}}
\end{figure}
\begin{figure}[t]
 \includegraphics[width=.80 \columnwidth]{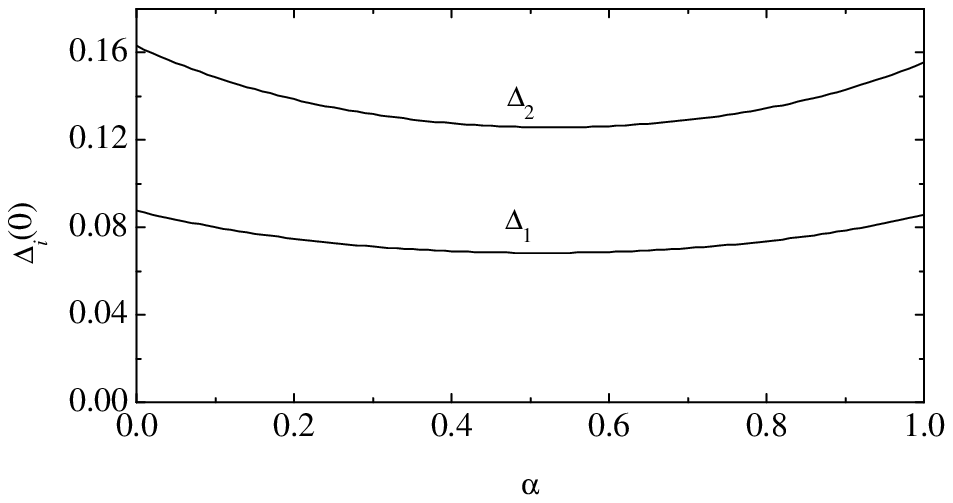}
\caption{Plot of the normal stress differences $\Delta_{1}=(P_{zz}-P_{xx})/{p}$ and $\Delta_{2}=(P_{zz}-P_{yy})/{p}$, evaluated at $y=0$, versus $\alpha$ for $g\lambda_0/v_0^2=0.05$.
\label{Delta0}}
\end{figure}

Equations (\ref{61}), (\ref{66}), and (\ref{67}) show that normal stress differences appear to order $g^2$. It is easy to check that $P_{yy}<P_{xx}<p<P_{zz}$, i.e.,  normal stresses are maximal along the flow direction and minimal along the direction normal to the plates. 
In order to characterize the normal stress differences, let us define the viscometric quantities
\beq
\Delta_{1}(y)\equiv 
\frac{P_{zz}(y)-P_{xx}(y)}{p(y)},\quad \Delta_{2}(y)\equiv 
\frac{P_{zz}(y)-P_{yy}}{p(y)}.
\label{72}
\eeq
Their expressions are
\beqa
\Delta_1(y)&=&\left[150\pi \frac{827-733\alpha+158\alpha^2}{(1+\alpha)^2(23-11\alpha)^2(3-\alpha)(7-4\alpha)}\right.
\nn
&&\left. +8\left(\frac{y}{\lambda_0}\right)^2\right]\left(\frac{g\lambda_0}{v_0^2}\right)^2+\mathcal{O}(g^4),
\label{4.6}
\eeqa
\beqa
\Delta_2(y)&=&\left[6\pi \frac{38467-34763\alpha+7708\alpha^2}{(1+\alpha)^2(23-11\alpha)^2(3-\alpha)(7-4\alpha)}\right.
\nn
&&\left. +\frac{56}{5}\left(\frac{y}{\lambda_0}\right)^2\right]\left(\frac{g\lambda_0}{v_0^2}\right)^2+\mathcal{O}(g^4).
\label{4.7}
\eeqa
Figure \ref{Delta12} shows the profiles of $\Delta_1(y)$ and $\Delta_2(y)$ for $\alpha=0.5$, $\alpha=0.8$, and $\alpha=1$ in the case $g\lambda_0/v_0^2=0.05$. We observe that the normal stress differences increase with the separation from the mid layer $y=0$. Moreover, those differences are more important for elastic gases ($\alpha=1$) than for inelastic gases ($\alpha=0.8$ and $\alpha=0.5$). However, as in the case of  the quantities plotted in Fig.\ \ref{ymaxTmax}, the $\alpha$-dependence of $\Delta_1$ and $\Delta_2$ is not monotonic. This is illustrated in Fig.\ \ref{Delta0} for the point $y=0$. We observe that the minimum values of $\Delta_1(0)$ and $\Delta_2(0)$ occur at $\alpha\approx 0.5$. 

\section{Concluding remarks\label{sec5}}
In this paper we have carried out a kinetic theory study of the steady planar Poiseuille flow undergone by a dilute granular gas under the action of the acceleration of gravity. In order to compensate locally for the energy loss due to the inelasticity of collisions, an external energy input in the form of a white noise driving  has been assumed. This type of driving mechanism has been introduced in the literature to mimic the heating effects due vibrating boundaries without the complications associated with boundary effects. This is especially convenient in our approach, since we have been mainly interested in the \textit{bulk} properties of the gas, namely in a slab centered in the middle layer having a width of the order of several mean free paths, away from the walls. 

Since granular gases are made of \textit{mesoscopic} particles, terrestrial gravity ($g=9.8~\text{m/s$^2$}$) plays in general a relevant role, in contrast to the case of molecular gases. The dimensionless parameter characterizing the influence of gravity during the free flight of a particle between two successive collisions is $g\lambda/v_{\text{th}}^2$, where $\lambda$ is the mean free path and $v_{\text{th}}$ is a typical (thermal) velocity. 
Under many conditions of practical interest, the parameter $g\lambda/v_{\text{th}}^2$ can have a non-negligible effect and yet be sufficiently small as to  justify a perturbative treatment. For instance, if $\lambda\sim 1~\text{mm}\text{--}1~\text{cm}$ and  $v_{\text{th}}\gtrsim 1~\text{m/s}$, which are typical values in
experiments on metallic or glass spheres, one can have $g\lambda/v_{\text{th}}^2\sim 10^{-3}\text{--}10^{-1}$.
Therefore, in our study we have performed a perturbation expansion of the velocity distribution function in powers of the gravity strength through second order. The reference state (i.e., the state at zero gravity) is the steady uniform state heated by a white noise thermostat. Since the Boltzmann equation for inelastic spheres is quite complicated to deal with, we have employed a kinetic model equation inspired in the  BGK model. This has allowed us to obtain explicitly the velocity distribution function through second order in terms of the velocity vector, the spatial coordinate, and the coefficient of restitution. By velocity integration one can obtain any desired moment, but here we have focused on the hydrodynamic fields (pressure, flow velocity, and granular temperature) and their associated fluxes (stress tensor and heat flux vector).

The results show that the non-Newtonian features previously studied in the case of elastic particles\cite{TS94,TSS98,TS01,STS03,RC98,HM99,ATN02} persist when inelasticity is present. In particular, the temperature profile $T(y)$ exhibits a bimodal shape: it has a local minimum $T_0$ at the central layer and reaches two symmetric maxima $T_{\text{max}}$ at a distance $|y_{\text{max}}|$ of about three mean free paths. The relative height of the two maxima, $(T_{\text{max}}-T_0)/T_0$ is about 10 times the square of the dimensionless parameter $g\lambda/v_{\text{th}}^2$. On the other hand, the heat flows outward from the central layer, so it goes from the colder to the hotter layers within the  region $|y|<|y_{\text{max}}|$. Other non-Newtonian effects include normal stress differences and the existence of a component of the heat flux parallel to the flow and hence normal to the thermal gradient.

The fact that the nonlinear transport properties of the granular Poiseuille flow are qualitatively similar to those of the elastic case does not come as a surprise, especially since the characteristic collisional cooling of the granular gas is balanced by an external driving. In that context, our aim in the present work has been two-fold. On the one hand, the example of gravity-driven Poiseuille flow allows one to emphasize once more that granular gases constitute an excellent playground to reveal interesting (and even counter-intuitive) non-Newtonian phenomena on scales accessible to laboratory conditions. More importantly, we wanted to assess the influence of inelasticity on the departure of the Poiseuille profiles from the Navier--Stokes predictions. This influence is not easy to foretell \textit{a priori} by means of intuitive or hand-waving arguments. According to the results reported in this paper, for small or moderate inelasticity (say $\alpha\gtrsim 0.5$) there is a slight decrease in the  quantitative deviations from the Navier--Stokes profiles as inelasticity grows: the  two temperature maxima  becomes lower and closer, while  the normal stress differences become smaller. The opposite behavior takes place for high inelasticity ($\alpha\lesssim 0.5$), although  that range is less interesting from an experimental point of view.

\begin{acknowledgments}   
A.S. is grateful to J. W. Dufty for discussions about the topic of this paper. The research of A.S. has been partially supported by the Ministerio de   
Ciencia y Tecnolog\'{\i}a   
 (Spain) through grant No.\ FIS2004-01399.
\end{acknowledgments}

\appendix
\section{Expressions for the coefficients \lowercase{$b_i$},
\lowercase{$c_i$}, and \lowercase{$d_i$}\label{appA}}

In this Appendix we list the explicit expressions of the coefficients in the expression for the velocity distribution function to order $g^2$, Eq.\ (\ref{56}). They are
\beq
b_0=-\frac{4}{5}\frac{(2+5\z)(5+7{\z}+6{\z}^2)}{(1+\z)^2(2+\z)(1+2\z)(2+3\z)},
\label{A1}
\eeq
\beq
b_1=\frac{8}{25}\frac{160+622\z-1051{\z}^2-2829{\z}^3-1696{\z}^4-276{\z}^5}
{(1+\z)^2(2+\z)^2(2+7\z+6{\z}^2)},
\label{A2}
\eeq
\beq
b_2=-\frac{48}{5}\frac{1+3\z}{(1+\z)(2+\z)(2+3\z)},\quad b_3=0,
\label{A3}
\eeq
\beq
b_4 = 
    -\frac{32}{5}\frac{5 + 29 \z + 12 {\z}^2)}{(1 + \z) (2 + \z) (1 + 2 \z)
    (2 + 3 \z)},
\label{A4}
\eeq
\beq
b_5 = 
    \frac{32}{5}\frac{(5 + 2\z)(1 + 3\z)}{(1 + \z)(2 + \z)(2 + 3\z)}, \quad b_6 = -\frac{16}{5}\frac{{\z}}{1 + \z},
\quad b_7 = -\frac{8}{3},
\eeq
\beq
c_0 = 
    8\frac{4 + 12\z + 113{\z}^2 + 176{\z}^3 + 
            79{\z}^4 + 6{\z}^5}{(1 + \z)^2
        (2 + \z)^2(1 + 2\z)(2 + 3\z)},
\label{A8}
\eeq
\beq
c_1 = 
    -32\frac{(3 + 2\z)
        (2 - 11\z - 12{\z}^2)}{(1 + \z)^2
          (2 + \z)^2(1 + 2\z)
        (2 + 3\z)},
\label{A9}
\eeq
\beq
c_2 = 
    16\frac{ 8 - 22\z - 23{\z}^2 - 3{\z}^3}{(1 + \z)(2 + \z)^2
            (2 + 3\z)}, \quad c_3 = 8\frac{{\z}}{1 + \z},
\label{A10}
\eeq
\beq
c_4 = 
    64\frac{3 + 2\z}{(1 + \z)(2 + \z)
        (1 + 2\z)(2 + 3\z)},
\label{A12}
\eeq
\beq
c_5 = -64\frac{3 + 2\z}{(1 + \z)(2 + \z)(2 + 3\z)}, \quad c_6 = \frac{16}{1 + {\z}},
\label{A13}
\eeq
\beq
d_0 = 
    \frac{8}{25}{\z}\frac{ 
        76 - 48\z - 137{\z}^2 + 7{\z}^3 + 
            42{\z}^4}{(1 + \z)^2
        (2 + \z)^2(1 + 2\z)(2 + 3\z)},
\label{A15}
\eeq
\beq
d_1 = 
    \frac{16}{25}\frac{76 - 48\z - 137{\z}^2 + 7{\z}^3 + 
            42{\z}^4}{(1 + \z)^2
        (2 + \z)^2(1 + 2 \z)(2 + 3\z)},
\label{A16}
\eeq
\beq
d_2 = 
    -\frac{16}{25}\frac{76 + 40\z + 23{\z}^2 + 21{\z}^3}{(1 + \z)(2 + \z)^2
            (2 + 3\z)}, \quad d_3 = -\frac{8}{5}\frac{{\z}}{1 + \z},
\label{A17}
\eeq
\beq
d_4 = 
    -\frac{64}{5}\frac{1}{(1 + \z)(1 + 2\z)
            (2 + 3\z)}, 
\label{A19}
\eeq
\beq
d_5 = \frac{64}{5}\frac{1}{(1 + \z)(2 + 3\z)}, \quad d_6 = -\frac{16}{5}\frac{1}{1 + \z},
\quad d_7 = \frac{16}{15}.
\label{A20}
\eeq

\end{document}